# Statistical methods for assessing non-replicable, outlying, and influential studies

Yefeng Yang[a,b,*], Shinichi Nakagawa[b,c]

[a]AgriAGI Research, Department of Biosystems Engineering, Zhejiang University, Hangzhou, 310058, China

[b]E&ERC, School of Biological, Earth and Environmental Sciences, University of New South Wales, Sydney, NSW 2052, Australia

[c]Department of Biological Sciences, University of Alberta, CW 405, Biological Sciences Building, Edmonton, AB T6G 2E9, Canada

[*]Corresponding author

yefeng.yang@zju.edu.cn/yefeng.yang1@unsw.edu.au

**Abstract**

Quantitative evidence synthesis (e.g., meta-analysis methods) has become a central tool for integration of findings across multiple studies, multi-centre trials, and multi-source cohort data. However, the identification and interpretation of non-replicable, outlying, and influential studies remain insufficiently addressed in practice, despite their potential to substantially affect the robustness and credibility of meta-analytic conclusions. In this paper, we clarify the conceptual distinctions between non-replicability, statistical outlyingness, and study influence, emphasizing that these concepts are related but not interchangeable. We then review the standard principles and procedures of model diagnostics for detecting outlying and influential studies in meta-analysis, together with their underlying statistical rationale. Building on recent methodological developments, we further discuss several practical and methodological refinements, including approaches for handling imprecise and correlated sampling variances, robust diagnostic procedures, and graphical tools for facilitating the identification and interpretation of unusual studies. Finally, we summarize recent advances in outlier and influence diagnostics and provide recommendations for the cautious interpretation and evaluation of studies identified as potentially non-replicable, outlying, or influential within meta-analytic frameworks.

## 1. Introduction

Achieving replicability is central to the cumulative progress of science. Nevertheless, large-scale preregistered multi-laboratory replication projects have revealed unexpectedly low rates of successful replication in several disciplines, contributing to what has become known as the replication crisis. For example, the Reproducibility Project: Cancer Biology sought to evaluate the reproducibility of preclinical cancer biology by independently repeating selected experiments from 53 influential papers published in journals such as *Nature*, *Science*, and *Cell* (Errington et al. (2014)). The project reported a successful replication rate of only 46% (Errington, Mathur, et al. (2021)). Although further empirical assessments are needed to determine the prevalence of non-replicability across disciplines, large-scale replication projects remain expensive, time-consuming, and often infeasible for many published findings (Hedges and Schauer (2019)). Indeed, only 50 of the 193 initially planned experiments in the cancer biology project were ultimately completed, largely because of incomplete methodological reporting, limited access to original data and materials, and practical difficulties encountered during replication attempts (Errington, Denis, et al. (2021)).

In parallel with these empirical efforts, a growing statistical literature has developed computational approaches for evaluating replicability in the absence of direct replication studies. One example is the z-curve framework, which estimates the expected replication rate by modeling the distribution of statistically significant results and the corresponding average statistical power(Brunner and Schimmack (2020), Bartoš and Schimmack (2022)). More recently, Xiao et al. (2024) have recently introduced two measures, the externally standardized residual ($r_{i(-A_{m,k})}$; Eq.

3.4) and its counterpart $R_{(A_{m,k})}$, derived from a leave-m-studies-out procedure to identify potentially non-replicable studies within a meta-analysis. Their work addresses an important problem; however, several aspects of the proposed framework warrant further clarification, including the interpretation of these measures, the distinction between statistical outlyingness and non-replicability, certain modelling assumptions, and the design of the simulation study used to evaluate the procedure.

The objective of the present paper is threefold. First, we examine several conceptual and methodological aspects of Xiao et al. (2024) that may benefit from further clarification. Second, motivated by these issues, we review the mathematical foundations of standard diagnostic procedures for identifying outlying and influential studies in meta-analysis and discuss several refinements that incorporate recent methodological developments. Third, we summarize recent advances in model-diagnostic methodology and conclude with practical recommendations for the interpretation and treatment of studies that appear unusual within a meta-analytic dataset. All illustrative R code used in this paper is available at the project repository: https://yefeng0920.github.io/AOAS_commentary/.

## 2. Conceptual and methodological considerations arising from Xiao et al. (2024)

The framework proposed by Xiao et al. (2024) addresses an important and timely problem, namely the statistical identification of studies that may not be replicable within a meta-analytic setting. Their work also highlights the broader need for clearer connections between methods for assessing non-replicability and established diagnostic procedures for detecting outlying and influential studies. In this section, we discuss several conceptual and methodological aspects of their framework that may benefit from further clarification. Our intention is not to dismiss the contribution of Xiao et al. (2024), but rather to place their proposal within the existing methodological literature and to clarify several distinctions that are important for the interpretation of diagnostic statistics in meta-analysis.

First, it is important to clarify that the externally standardized residual $r_{i(-A_{m,k})}$, which Xiao et al. (2024) claim as a novel measure (see their Abstract), is not genuinely new. The externally standardized residual $r_{i(-A_{m,k})}$ was first introduced in the context of meta-analysis over 15 years ago by Viechtbauer and Cheung (2010) (see their Eqs. 15 and 16). It represents a natural extension of the externally standardized residuals developed for normal linear regression (Ling (1984), Hedges and Olkin (1985)). Regrettably, Xiao et al. (2024) failed to cite these foundational works, thereby overlooking the significant contributions by Viechtbauer and Cheung (2010) and Hedges and Olkin (1985). Additionally, their Eqs. 3.2 and 3.3 are not new innovations, as these expressions have been previously proposed and extended in the meta-analysis literature, particularly in the context of mean-shift models designed to detect outlier studies that contribute extreme effect sizes (Negeri and Beyene (2020), Noma et al. (2020), Metelli et al. (2023); see also "Section 4 Recent advances in outlying and influential studies detection techniques"). We contend that that when extending or adapting established statistical methods, it is crucial to acknowledge prior work appropriately. Proper citation of relevant literature is a fundamental principle of rigorous scientific practice.

Second, while $r_{i(-A_{m,k})}$ was initially introduced as a diagnostic measure to identify outlying studies, Xiao et al. (2024), drawing on the work by Schauer and Hedges (2020), derived a variate $R_{(A_{m,k})}$, which is the sum of squared $r_{i(-A_{m,k})}$, applied it 'innovatively' to quantify non-replicable studies within a meta-analysis. This application implicitly operationalizes non-replicable studies as studies that behave as statistical outliers relative to the assumed meta-analytic model. While this perspective is reasonable within a model-diagnostic framework, it represents only one possible conception of replicability. Replicability has been defined in different ways across the literature (Hedges and Schauer (2019)), including consistency in effect direction, magnitude, statistical significance, or reproducibility of inference. Consequently, outlyingness and non-replicability should not be regarded as interchangeable concepts, even though they may overlap in some situations. A study may be statistically outlying while still being scientifically replicable under certain definitions, whereas studies with inconsistent or unstable findings may not necessarily appear as statistical outliers within a meta-analysis. The more commonly adopted operational definition of replicability involves observing a statistically significant replication effect in the same direction as the original effect (e.g., a two-sided p-value $< 0.05$) in the same direction as the original effect (e.g., Jaljuli et al. (2023)), a criterion employed in most large-scale replication efforts (Collaboration (2015), Camerer et al. (2016), Errington, Mathur, et al. (2021)). For instance, in an examination of 18 laboratory experiments in economics, 11 replication studies (61%) reported statistically significant effects in the same direction as the original studies (Camerer et al. (2016)). Another common operational definition involves comparing the effect size from the original study with the 95% confidence interval of the effect size from the replication studies to determine whether there is a significant difference.

It is important to clarify that in many meta-analyses, certain studies may exhibit markedly different characteristics (e.g., population, intervention, outcomes) from others, resulting in either unusually large or small effect size estimates, thereby classifying them as outliers. For these outlying studies, we advocate for a thorough examination of their characteristics (Hedges (1986), Viechtbauer and Cheung (2010)) rather than arbitrarily assuming they are inherently non-replicable studies, as Xiao et al. (2024) did. Outlying studies, particularly those with distinct characteristics, may point to new hypotheses, such as potential moderators that could explain the extreme effect size estimates. It is also plausible that outliers are artifacts, potentially arising from errors such as the miscalculation of effect sizes (e.g., using the standard error instead of the standard deviation in computing Cohen's d), inappropriate study selection criteria, low study quality, or selective reporting (Viechtbauer (2010), Pereira et al. (2012), Desai et al. (2019)). Although there are no definitive guidelines for handling outlying studies when they are detected, treating them as synonymous with non-replicable studies is unjustifiable. A non-replicable study, by contrast, suggests that the original study's effect size was likely a false positive; noise misinterpreted as a signal. It might also indicate that the original finding was overly generalized and is only replicable under more specific conditions than initially thought. Furthermore, non-replicability may imply that the methodologies required to achieve the observed effect size are not sufficiently well-defined or understood (Errington, Denis, et al. (2021)), or that the theoretical explanation for the finding is flawed. Additionally, failures in replicating experimental protocols could contribute to non-replicability (Errington, Mathur, et al. (2021)).

Third, while Xiao et al. (2024) proposed using a leave-$m$-studies-out procedure to calculate externally standardized residual $r_{i(-A_{m,k})}$ and thus $R_{(A_{m,k})}$, their simulation studies, software implementation, and case studies are limited to examining the case where $m = 1$. Consequently, the empirical performance of their proposed measure for values of $m$ other than 1 remains untested. When $m = 1$, the leave-$m$-studies-out procedure effectively reduces to a leave-one-out analysis, a conventional diagnostic tool that identifies outliers by refitting the specified meta-analysis model while sequentially omitting each study (see "Section 3. Standard practice of outlying and influential studies detection"; Viechtbauer (2010)). For example, the *leave1out()* function in the metafor R package facilitates an accessible implementation of the leave-one-out analysis in meta-analysis (Viechtbauer (2010)). Figure 1, generated using leave1out(), demonstrates that omitting study 15 (MASS III 2009 by Xiao et al. (2024)) markedly shifts the overall mean effect size, between-study variance, and heterogeneity, suggesting it is a likely outlier, consistent with the findings of Xiao et al. (2024). However, it is important to point out the theoretical limitations of the leave-$m$-studies-out procedure for detecting outliers. Specifically, the presence of multiple outliers can result in misleading diagnostics, either by masking, where no true outliers are detected, or by swamping, where some studies are incorrectly identified as outliers (Barnett and Lewis (1994), Viechtbauer (2020)).

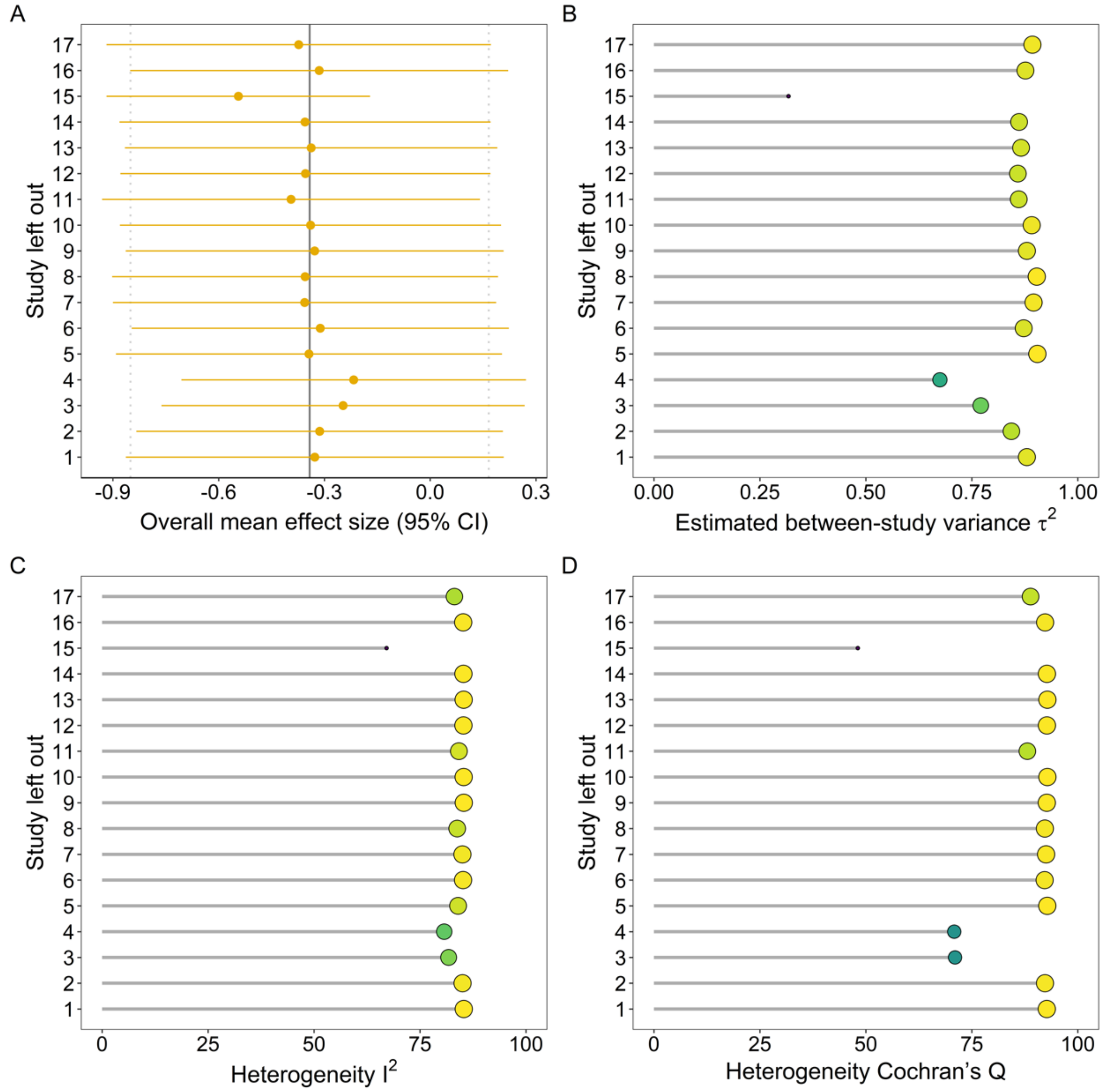

Figure 1 Leave-one-out analysis in meta-analysis based on the data from case studies in Xiao et al. (2024). (A) The estimated overall average effect size (or, grand mean) derived from leaving out a study from the model. (B) The estimated between-study variance $\tau^2$ derived from leaving out a study from the model. (C) The amount of heterogeneity $I^2$ derived from leaving out a study from the model. (D) Cochran's test statistics for the test of heterogeneity derived from leaving out a study from the model. All of the results corresponding to the leave-one-out analysis indicate that study 15 (MASS III 2009 by Xiao et al. (2024)) is a potential outlying study. 95% CI denotes the 95% confidence interval.

Fourth, several aspects of the formulation in Xiao et al. (2024) warrant further clarification. First, the residual quantity introduced in their subsection 3.3 appears to correspond to the raw residual divided by the square root of the model-implied marginal variance. In the meta-analytic literature, this quantity is directly adapted from the standard linear model (Cook and Weisberg (1982)), where it is commonly described as a *Pearson residual* or *semi-standardized residual*, rather than the internally standardized residual, because the denominator reflects the marginal variance $v_i + \hat{\tau}^2$ instead of the estimated sampling variance of the residual itself. Under this formulation, the squared residuals contribute directly to Cochran's $Q$ statistic (Cochran (1954), and therefore describe departure from the fitted model in a different way from studentized residuals used for formal outlier assessment. Although the numerical expression is familiar, distinguishing among these residual definitions is important because different residuals serve different diagnostic purposes in practice.

A related consideration concerns the variance structure assumed in Equations 3.2 and 3.3 of Xiao et al. (2024). In those equations, studies classified as non-replicable are modelled as differing from the remaining studies through a shift in their expected effect, while the between-study variance parameter $\tau^2$is assumed to remain common across both groups. Such a specification is mathematically admissible and corresponds to a location-shift formulation in which atypical studies differ in mean rather than dispersion. Nevertheless, in practical applications the presence of extreme studies often inflates the *estimated* heterogeneity, $\hat{\tau}^2$, when all studies are analyzed jointly under a conventional random-effects model (Viechtbauer and Cheung (2010)). Because residual-based diagnostics depend directly on this estimated variance component, an inflated $\hat{\tau}^2$ can enlarge the denominator of residual statistics and thereby reduce sensitivity for detecting unusual studies, as illustrated in Figure 1 based on the example data from their case studies. This masking phenomenon has been noted in earlier work and was one of the motivations for deleted-case diagnostics that recompute the variance component after omitting the candidate study. From this perspective, the assumption of a common variance is not itself problematic, but its implications for residual-based detection merit explicit discussion when distinguishing non-replicable studies from conventional outliers. In network meta-analysis, outlying studies can contribute as a major source of heterogeneity (Noma et al. (2020), Petropoulou et al. (2021), Metelli et al. (2023)). A likelihood-ratio test has even been developed to determine whether the random effect variances specific to certain effect size estimates are inflated, identifying such studies as outliers when this is the case (Gumedze and Jackson (2011)).

Fifth, a further point concerns the parameterization of their Equation 3.3. As presented, the model introduces an additional location parameter for each study classified within the candidate non-replicable subset. When the subset contains several studies, this formulation substantially increases the number of parameters that must be estimated from a comparatively small amount of information. Under such circumstances, the model may become weakly identified, so that the individual shift parameters cannot be estimated with adequate precision and the resulting inference becomes sensitive to small perturbations in the data. In practice, models of this type usually require additional structural assumptions, shrinkage constraints, or alternative parameterizations to ensure stable estimation. For this reason, in the subsequent section we restate the corresponding formulation within the established mean-shift framework for meta-analysis, which yields a more transparent interpretation of study-specific deviations while preserving statistical identifiability.

Two aspects of description of $r_{i(-A_{m,k})}$ would benefit from clarification. First, under the notation of the original paper, a separate shift parameter is introduced for each study contained in the candidate subset $A$, so that the number of additional parameters increases with the size of that subset, $|A|$. When $|A|$ is not small, the amount of information available for estimating these study-specific deviations may become limited relative to the complexity of the model. As a consequence, the resulting system can be weakly identified, and the estimated shifts may become unstable or highly sensitive to small changes in the observed data. This issue does not invalidate the underlying formulation, but it does indicate that estimation may require additional structure, such as shrinkage or alternative parameterizations, to ensure numerical stability. In the subsequent section, we restate the corresponding model within the established mean-shift framework used in meta-analysis, which provides a more transparent representation of study-specific departures while preserving interpretability.

$r_{i(-A_{m,k})}$ may itself be viewed as a Wald-type diagnostic statistic for assessing whether a given study is unusually inconsistent with the fitted model. Specifically, the numerator corresponds to the deleted residual obtained after omitting the $i$-th study, whereas the denominator is the estimated standard error of that deleted residual, yielding the usual studentized form described by Viechtbauer and Cheung (2010). Under the assumed model, this quantity can therefore be interpreted directly as a test statistic for evaluating departure from the postulated sampling structure. Although Xiao et al. (2024) further introduced an aggregated statistic formed by averaging the $r_{i(-A_{m,k})}$ values across the $m$ omitted studies, the manuscript does not provide a formal derivation or comparative evidence demonstrating that this summary statistic offers advantages over the study-specific externally studentized residuals on which it is based.

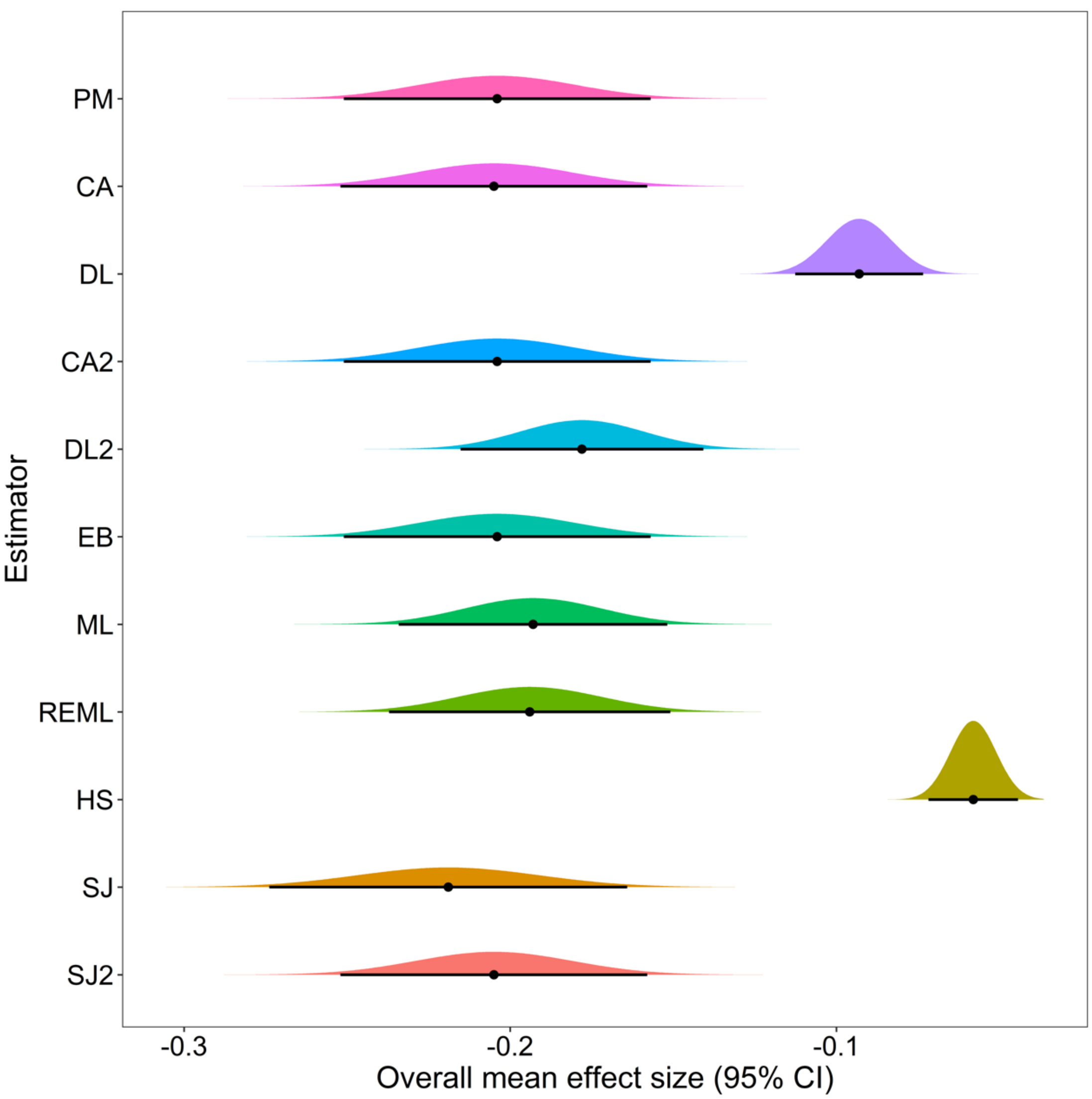

Figure 2. The sensitivity analysis examining the robustness of the meta-analysis results to different estimators. A total of 11 estimators were used. Results for the overall mean effect size are shown in the figure. Additional results can be added to the figure, including between-study variance $\tau^2$, heterogeneity ($I^2$), and Cochran's test statistics (see Figure 1). PM is the Paule-Mandel estimator.CA is the Cochran estimator (equivalent to the Hedges or variance component estimator). DL is the DerSimonian-Laird estimator (equivalent to the general constant method CA2 is a two-step Cochran estimator. DL is a two-step variance estimator. EB is empirical Bayesian estimator. ML is the maximum likelihood estimator. REML is the restricted maximum likelihood estimator. HS is the Hunter-Schmidt estimator. SJ is Sidik-Jonkman estimator. SJ2 is a modified the Sidik-Jonkman estimator, using the Cochran estimator. SJ2 is a modified Sidik-Jonkman estimator that uses the estimates from the Cochran estimator as the starting values. Empirical performance and mathematical details of these estimators can be found in DerSimonian and Kacker (2007) and Viechtbauer (2007). 95% CI denotes the 95% confidence interval.

Sixth, a further point concerns the description of the simulation design in Xiao et al. (2024). In subsection 3.2, both replicable and non-replicable studies are described as sharing a common between-study variance parameter $\tau^2$. By contrast, the first paragraph of subsection 4.1 appears to describe the non-replicable studies as having no between-study heterogeneity, whereas the notation in the subsequent paragraph again suggests a common variance structure across both groups. This discrepancy may reflect a modelling problem or just a typographical inconsistency in the description of the simulation scheme. Nevertheless, because the simulation code was not made publicly available, it is difficult to determine which data-generating mechanism was actually implemented. We therefore believe that clearer reporting of simulation algorithms, together with accessible code, would improve the transparency and reproducibility of methodological studies of this kind. Thirdly, the authors utilized an extremely large, and arguably unrealistic, effect size for non-replicable studies: 3 (in log odds ratio). A log odds ratio of 3 implies that the odds of the event occurring in the exposed group are approximately 20 times ($e^3$) times higher than in the unexposed group. However, in a review of 171 epidemiological studies, the median odds ratio was 2.16 (Chen et al. (2010)). Well-cited "rules of thumb" suggest that odds ratios of 1.68, 3.47, and 6.71 correspond to Cohen's d values of 0.2 (small), 0.5 (medium), and 0.8 (large), respectively (Chen et al. (2010)). Therefore, in practical terms, the odds ratio of 20 used by Xiao et al. (2024) represents an extremely large effect size, indicating an unusually strong association between exposure and outcome.

Seventh, Xiao et al. (2024) also raise an important practical question regarding the extent to which atypical studies may affect meta-analytic conclusions. In this context, it is useful to distinguish between outlyingness and influence, because these concepts describe different features of a study within a meta-analysis. An outlying study is one whose observed effect differs substantially from the pattern predicted by the fitted model, whereas an influential study is one whose inclusion materially changes parameter estimates, uncertainty, or inferential conclusions. Although some outlying studies may also be influential, the two properties do not necessarily coincide. A study can appear statistically unusual yet have little impact on the pooled estimate, whereas a study with a moderate residual discrepancy may substantially affect inference because of its weight or leverage. For this reason, studies identified as outliers should not automatically be interpreted as capable of reversing the substantive conclusions of a meta-analysis unless their influence on model estimates is formally evaluated. On the contrary, if a study is not identified as an outlier, it could still challenge the conclusions due to unusual characteristics like population, intervention, outcome, or study design, as illustrated in case studies by Viechtbauer and Cheung (2010) and Viechtbauer (2020). We further clarify that Xiao et al. (2024) incorrectly claim that the restricted maximum likelihood (REML) estimator has the best performance in most settings. The reality, as demonstrated by the extensive simulation in Langan et al. (2019) is that none of the nine heterogeneity estimators consistently performs well, particularly in meta-analyses with a small number of studies or limited within-study sample sizes. While the two-step DerSimonian-Laird estimator and REML generally offer the most reasonable performance for meta-analyses using standardized mean difference and odds ratio, REML is already widely known and implemented in most statistical software packages. As a result, it is recommended. Nonetheless, methodologists advocate for conducting sensitivity analyses with a variety of estimators (ideally at least two to three) to evaluate the robustness of conclusions, especially in meta-analyses with a small number of studies (Veroniki et al. (2016)). Using publicly available data (Nisa et al. (2019)), we demonstrated that the conclusions of a meta-analysis can shift significantly depending on the chosen estimator (Figure 2).

Although the discussion in this paper is framed within the frequentist meta-analytic framework, many of the conceptual issues discussed here extend naturally to Bayesian meta-analysis. In particular, the distinction between non-replicable studies, outlying studies, and influential studies remains equally important under Bayesian models. Bayesian approaches do not eliminate the need for model diagnostics; rather, they provide additional diagnostic tools, such as posterior predictive checks, leave-one-out cross-validation, and influence measures based on posterior distributions. Similarly, outlying studies may still exert substantial influence on posterior estimates, especially when the number of studies is small or prior information is weak. Future work could explore how recent advances in Bayesian model diagnostics may complement the methods reviewed here.

## 3. Standard practice of outlying and influential studies detection

The objectives of this section are threefold. First, we aim to reformulate the methods introduced by Viechtbauer and Cheung (2010) as the standard approach for detecting outlying and influential studies in meta-analysis. To emphasise the mathematical foundations of these methods, we will briefly restate them. Implementing these methods is readily accessible via the R package metafor (Viechtbauer (2010)). Second, we offer three refinements to the methods developed by Viechtbauer and Cheung (2010), to address typical violations of model assumptions. Third, we introduce visual approaches (illustrated in Figure 3) that effectively present the results generated by the metafor package to aid the detection of outlying and influential studies. We begin with an overview of widely used statistical models for meta-analysis.

### 3.1. Unified statistical model for meta-analysis

Consider a meta-analysis involving $k$ independent studies, each providing a single effect size estimate. The outcome measure may vary depending on the meta-analysis's objectives and the statistical properties of the chosen effect size measures (Viechtbauer (2010)). For instance, Cohen’s $d$ (or its variant, Hedge’ s $g$) is typically employed when assessing the effectiveness of an intervention. Alternatively, log-transformed odds ratios (or variants like the risk ratio) are used when the studies present results in terms of 2-by-2 contingency tables with dichotomous outcomes. Regardless of the outcome measure used, let $y_i$ denote the effect size estimate (observed effect), and $\theta_i$ represent the corresponding effect size parameter (true effect) for the $i$-th study. The following notation adopts the conventional hierarchical formulation of meta-analysis models described in standard methodological texts (Hedges and Olkin (1985), Viechtbauer (2010)). In the standard meta-analysis framework, the effect size estimate $y_i$ is modelled as a random draw from a normal distribution (Hedges and Olkin (1985)):

$$y_i = \theta_i + e_i, (1)$$

where $e_i$ represents the within-study sampling error with distribution $N(0, v_i)$, and $v_i$ is the estimated within-study sampling variance associated with $y_i$. T The assumption of independence among the $k$ studies implies that the sampling errors are mutually uncorrelated, such that $\text{Cov}(e_i, e_{i'}) = 0$ for $i \neq i'$. When this assumption is violated, for example because multiple outcomes are measured on the same participants, modifications to this framework are required and are discussed below.

The normal sampling model provides two practical advantages. First, the observed effect size $y_i$is an unbiased estimator of the underlying study-specific effect $\theta_i$. Second, sampling theory permits analytic approximations for $v_i$, often obtained through large-sample expansions such as the delta method. In routine applications, the estimated sampling variance $v_i$is commonly treated as fixed when constructing inverse-variance weights, thereby determining the relative precision assigned to each study. Although this approximation simplifies statistical inference, it does not account for uncertainty in the estimation of $v_i$. Several methodological extensions have therefore been developed to incorporate this additional source of uncertainty more explicitly (Doncaster and Spake (2018), Veroniki et al. (2019)).

Although the true data-generation process of the parameter $\theta_i$ is typically unknown, it can often be represented as a linear combination of predictor variables plus a random error term, indicating potential heterogeneity in $\theta_i$:

$$\theta_i = \beta_0 + \beta_1 x_{i1} + \beta_1 x_{i1} + \cdots + \beta_p x_{ip} + u_i, (2)$$

where $x_{ij}$ denotes the value of the $j$-th moderator variable (explanatory variable or independent variable in the framework of standard linear regression), $\beta_j$ ($j = 1, \ldots, p$) represents the corresponding model coefficient (slope) indicating the change in the effect size as $x_{ij}$ increases by one unit, and $\beta_0$ denotes the model intercept, $u_i$ denotes the study-specific random deviation from the linear predictor and is assumed to follow a normal distribution $N(0, \tau^2)$, where $\tau^2$ represents the residual between-study heterogeneity not explained by the moderator variables $x_{ij}$.

The average (expectation) of the true parameter $\theta_i$ across $k$ independent studies is termed the average true effect size (also known as the overall mean effect or grand mean):

$$E(\theta_i) = \beta_0 + \beta_1 x_{i1} + \cdots + \beta_p x_{ip} = \mu_i, \qquad (3)$$

where $\mu_i$denotes the study-specific mean implied by the moderator structure. Combining Equations (1) and (2) yields the general mixed-effects meta-regression model:

$$y_i = \beta_0 + \beta_1 x_{i1} + \cdots + \beta_p x_{ip} + u_i + e_i, \qquad (4)$$

or equivalently,

$$y_i = \mu_i + u_i + e_i, \qquad (5)$$

where $\mu_i = x_i^{\top}\beta$, $u_i \sim N(0, \tau^2)$ represents the study-specific random deviation and $e_i \sim N(0, v_i)$ denotes sampling error. Several familiar meta-analytic models arise as special cases of this formulation. When no moderator variables are included, $\mu_i \equiv \beta_0$, yielding the intercept-only random-effects model. When $\tau^2 = 0$, the model reduces to a fixed-effects meta-regression model, and when both no moderators are included and $\tau^2 = 0$, the common-effect model is obtained.

### 3.2. Step 1: Model fitting and inference

The first step in detecting outlying and influential studies involves fitting Eq. 4 or 5 (or their special cases). It is pertinent to note that when the diagnostic methods for model evaluation were first proposed by Viechtbauer and Cheung (2010), only a limited range of estimators were available. Since then, various estimators and enhanced inference methods have been developed. Thus, integrating these advancements into diagnostic methods is essential. In this section, we present three refinements that incorporate these methodological innovations. For clarity, we use matrix notation to streamline the mathematical expressions, while retaining algebraic notation where it aids comprehension. Using matrix notation, Eq. 4 can be compactly expressed as:

$$\boldsymbol{y} = \boldsymbol{X\beta} + \boldsymbol{u} + \boldsymbol{e}, (6)$$

where $\mathbf{y}$ is a $k \times 1$ column vector of effect size estimates, $\boldsymbol{X}$ denotes a $k \times (p + 1)$ model (design) matrix containing the values of the $p$ moderator variables a column of ones in the first column for the model intercept, $\boldsymbol{\beta}$ denotes a $(p + 1) \times 1$ column vector of model coefficients $\beta_0, \beta_1, \ldots, \beta_p$, and $\boldsymbol{u}$ and $\boldsymbol{e}$ denote $k \times 1$ vectors with $u_i$ and $e_i$ values. Eq. 6 can be reformulated in its marginal form $\boldsymbol{y} \sim N(\boldsymbol{X\beta}, \boldsymbol{M})$, where $\boldsymbol{M} = \tau^2 \boldsymbol{I} + \boldsymbol{V}$ is the so-called 'marginal' variance-covariance matrix of the effect size estimates, $\boldsymbol{V}$ is a $k \times k$ diagonal matrix with elements representing the sampling variances $v_i$, and $\boldsymbol{I}$ is a $k \times k$ identity matrix.

According to the Gauss–Markov theorem, the weighted least squares estimator of the model coefficients is given by:

$$\boldsymbol{\beta} = (\boldsymbol{X}'\boldsymbol{W}\boldsymbol{X})^{-1}\boldsymbol{X}'\boldsymbol{W}\boldsymbol{y}, (7)$$

where $\boldsymbol{W} = \boldsymbol{M}^{-1}$, representing a $k \times k$ diagonal weight matrix with elements $w_i = 1/(v_i + \tau^2)$. Therefore, if the true value of between-study variance ($\tau^2$) is known, Eq. 7 provides the uniformly minimum variance unbiased estimator (UMVUE) of the model coefficients $\boldsymbol{\beta}$. In practice, $\tau^2$ is unknown and has to be estimated from the data, denoted by $\hat{\tau}^2$. This estimate is then plugged into Eq. 7 to obtain the estimated model coefficient $\widehat{\boldsymbol{\beta}}$. To quantify the precision of the $\widehat{\boldsymbol{\beta}}$, we compute their variance-covariance matrix:

$$\mathrm{Var}[\widehat{\boldsymbol{\beta}}] = (\boldsymbol{X}'\widehat{\boldsymbol{W}}\boldsymbol{X})^{-1}, (8)$$

The diagonal elements of $\mathrm{Var}[\widehat{\boldsymbol{\beta}}]$ represent the sampling variances of the model coefficients (i.e., $\mathrm{Var}[\hat{\beta}_0], \mathrm{Var}[\hat{\beta}_1], \ldots, \mathrm{Var}[\hat{\beta}_p]$), and the square roots of these variances yield the standard errors (i.e., $\mathrm{SE}[\hat{\beta}_0]$, $\mathrm{SE}[\hat{\beta}_1], \ldots, \mathrm{SE}[\hat{\beta}_p]$). Model inference can be conducted by constructing Wald-type test statistics $z_j = {\hat{\beta}_j}/{\mathrm{SE}[\hat{\beta}_j]}$. Thus, $z_j$ can be compared to the critical values of the standard normal distribution for a given significance level (e.g., ±1.96 for a two-sided test at $\alpha$ = 0.05). The 95% confidence intervals for $\hat{\beta}_j$ are constructed as $\hat{\beta}_j + 1.96\mathrm{SE}[\hat{\beta}_j]$. It is important to note that in practice, estimates of $\tau^2$ as well as $v_i$ may be imprecise or biased, leading to underestimation or overestimation of $\mathrm{SE}[\hat{\beta}_j]$ and potentially distorting model inference (e.g., non-nominal Type I error rates). Given that $\mathrm{SE}[\hat{\beta}_j]$, $\tau^2$, and $v_i$ are important components of the model diagnostic measures (see below or Eqs. 15, 19, 26 in Viechtbauer and Cheung (2010)), inaccuracies can lead to spurious conclusions regarding outlying and influential studies.

### 3.2.1 Refinement 1: Guarding against biased estimators

In Viechtbauer and Cheung (2010), the model diagnostic methods were initially introduced using only the DerSimonian-Laird (DL) estimator for estimating the between-study variance $\tau^2$. The DL estimator is a method-of-moments estimator that relies on the inverse of the sampling variances $v_i$ of effect size estimates (DerSimonian and Laird (1986)). Although the DerSimonian–Laird estimator has attractive analytical simplicity and can be approximately unbiased under ideal assumptions (Viechtbauer (2005), Bhaumik et al. (2012)), extensive simulation studies have shown that it may yield systematically biased estimates of between-study variance in realistic meta-analytic settings, particularly when study sample sizes are small or within-study variances are imprecisely estimated (Sidik and Jonkman (2007), Panityakul et al. (2013), Langan et al. (2019), Veroniki et al. (2019)). This limitation arises because the method treats the within-study sampling variances as fixed, even though these quantities are themselves estimated and may contain substantial uncertainty in small samples. Fortunately, methodological advancements have introduced several alternative estimators, including both iterative and non-iterative methods (Viechtbauer (2005), Sidik and Jonkman (2007), Panityakul et al. (2013), Viechtbauer et al. (2015), Veroniki et al. (2019). However, recommendations regarding these estimators from various simulation studies are rarely consistent (Langan et al. (2019)). This inconsistency suggests that no single estimator universally outperforms others across all conditions. Additionally, there may be a tendency for authors to recommend their own newly proposed estimators, potentially due to confirmation bias, which may hinder consensus.

To address potential confirmation bias and ensure robustness in model diagnostics, we recommend employing multiple estimators for fitting models and detecting outlying and influential studies. Based on the latest simulation study with minimal conflict of interest (i.e., no new estimators proposed; Langan et al. (2019)), we advocate using the restricted maximum likelihood (REML) estimator as the primary method, supplemented by at least one additional estimator for sensitivity analysis. The REML estimator maximizes the restricted log-likelihood of the model parameters. Using matrix notation, the maximum restricted log-likelihood for $\tau^2$ is expressed as:

$$ll_{REML} = -\frac{k-p-1}{2} ln(2\pi) + \frac{1}{2}\ln(|\boldsymbol{X}'\boldsymbol{X}|) - \frac{1}{2}\ln(|\boldsymbol{M}|) - \frac{1}{2}\ln(|\boldsymbol{X}'\boldsymbol{W}\boldsymbol{X}|) - \frac{1}{2}(\mathbf{y} - \mathbf{X}\boldsymbol{\beta})'\boldsymbol{W}(\mathbf{y} - \mathbf{X}\boldsymbol{\beta}), (9)$$

Given that no closed-form solutions exist for $\hat{\tau}^2$ with maximum $ll_{REML}$, numerical techniques are necessary to obtain this estimate. Various optimization algorithms have been implemented in the metafor package (Viechtbauer (2010)). Among these, the Fisher scoring algorithm is generally recommended due to its insensitivity to initial values and its ability to achieve convergence efficiently (Harville (1977), Viechtbauer et al. (2015)). In the first step, closed-form estimators such as the Hedges estimator, which is the default of the metafor package (Viechtbauer (2010)), can be used to provide a starting value for $\hat{\tau}^2$. Then, an iterative process is used to find the $\hat{\tau}^2$ with maximum $ll_{REML}$. During each iteration, the current estimate $\hat{\tau}^2_{current}$ is updated to a new value $\hat{\tau}^2_{new}$ based on a factor $\Delta$, which incorporates Fisher information and $ll_{REML}$ of each step:

$$\hat{\tau}^2_{new} = \hat{\tau}^2_{current} + \Delta, (10)$$

The iteration process continues until $\Delta$ is smaller than a pre-defined threshold (e.g., $10^{-5}$ by default in the metafor package; Viechtbauer (2010)), indicating convergence of the fitted model and identification of the parameters. If convergence issues arise (which is likely in simulation studies), the threshold and the number of iterations can be adjusted to ensure robust fitting of the model.

**3.2.2 Refinement 2: Guarding against the imprecise between-study variance**

As previously mentioned, the estimation of the variance-covariance matrix of model coefficients (Eq. 8) and the corresponding standard errors using the DerSimonian-Laird estimate of between-study variance ($\tau^2$) as the true parameter, disregards the uncertainty in the estimation of $\tau^2$ (Viechtbauer and Cheung (2010)). To address this issue, the Hartung-Knapp-Sidik-Jonkman (HKSJ) adjustment has been proposed to account for the uncertainty in $\tau^2$ (Hartung (1999), Sidik and Jonkman (2002), Knapp and Hartung (2003)). The adjusted variance-covariance matrix is given by:

$$\mathrm{Var}\left[\widehat{\boldsymbol{\beta}}\right]_{hksj} = s^2(\boldsymbol{X}'\widehat{\boldsymbol{W}}\boldsymbol{X})^{-1}, (11)$$

where $s^2$ is the adjustment factor, computed as:

$$s^2 = \frac{(\mathbf{y} - \mathbf{X\beta})' \boldsymbol{W} (\mathbf{y} - \mathbf{X\beta})}{k - p - 1}, (12)$$

The intuitive rationale behind the Hartung-Knapp-Sidik-Jonkman adjustment is that it relaxes the assumption that the sampling variance $v_i$ of effect size estimate is known precisely without uncertainty. Instead, it assumes that $v_i$ is known only up to a proportionality constant, closely related to $s^2$. This adjustment also establishes a connection between meta-analysis models and standard linear models (Van Aert and Jackson (2019)). Specifically, meta-analysis with the Hartung-Knapp-Sidik-Jonkman adjustment is equivalent to an intercept-only weighted least squares regression model where the weights are the reciprocals of $\frac{1}{v_i + \tau^2}$ as the weight.

#### 3.2.3 Refinement 3: Guarding against biased, imprecise and correlated sampling variances of effect sizes

Three factors may contribute to biased and imprecise sampling variances $v_i$, which can in turn impact the detection of outlying and influential studies. First, some effect size measures rely on approximate normality of the sampling distribution of the estimated effect size, and departures from this approximation can reduce the accuracy of the corresponding variance formulas. Second, the formulas for computing sampling variances are typically derived from asymptotic results, which assume large sample sizes; these formulas may be less reliable for small samples. Third, violations of the assumption of independence among effect size estimates (e.g., such as when multiple measurements are taken from the same individuals) can also affect variance estimation.

When the usual normal-approximation framework may be inadequate for particular outcome types, several alternative meta-analytic methods are available that model the underlying data more directly. These include the Mantel-Haenszel method, Peto's method, generalized linear mixed-effects models (such as Poisson and logistic regression), and the non-central hypergeometric model (Mantel and Haenszel (1959), Yusuf et al. (1985), Robins et al. (1986), Bradburn et al. (2007), Stijnen et al. (2010)). Specifically, for data presented in 2×2 tables (e.g., odds ratios), the mixed-effects conditional logistic model, based on the non-central hypergeometric distribution, can be utilized. For dichotomous data, mixed-effects logistic regression models, often referred to as "binomial-normal models," are appropriate. For event count data, mixed-effects Poisson regression models, or "Poisson-normal models," can be applied. These models are available in the metafor package (Viechtbauer (2010)).

To address the second issue, where sampling variances of effect size estimates might be inaccurately computed due to small sample sizes, various adjusted formulas have been proposed for specific effect size measures (Berkey et al. (1995)). For instance, Doncaster and Spake (2018)) suggest an adjustment for Cohen's d and log response ratio that involves substituting the sample mean of each study with a cross-study sample mean—either a simple or weighted mean—in the formula used to compute sampling variance. The rationale behind this adjustment is that it assumes the sampling variance of the effect size estimate is not solely influenced by the study-specific sample size but is instead drawn from a common population variance across all studies. The idea is that when the population mean is homogeneous, the cross-study sample mean offers a more accurate estimate of the population mean than the sample means from individual studies. A similar adjustment has been proposed for log relative risk (Berkey et al. (1995)). This rationale has also been extended to address issues related to incomplete reporting of sampling variance (e.g., missing standard deviations) in meta-analyses. Nakagawa et al. (2023) proposed a data imputation method tailored to a log response ratio, which demonstrated superior performance compared to traditional data imputation techniques. By using the cross-study mean, the adjustment aims to improve the precision of the variance estimate and thus enhance the detection of outlying and influential studies.

For the third issue, we recommend using robust variance estimation methods to compute robust standard errors for model coefficients (Sidik and Jonkman (2006), Hedges et al. (2010)). These methods address the problem of ignored sampling correlation, which is often not reported in primary studies. Robust variance estimation methods allow for an arbitrary assumption of sampling correlation (e.g., 0.5 or 0.8) in a "working" model (Yang et al. (2022)). Despite the misspecification of the correlation structure due to the arbitrarily assumed sampling correlation, the resulting standard errors—known as cluster-robust standard errors—are valid. These standard errors are derived from products of regression residuals, providing accurate estimates of the variance-covariance structure when the number of studies is large. Additionally, methods for small-sample adjustments have been developed to address issues related to small sample sizes (Tipton (2015), Tipton and Pustejovsky (2015), Joshi et al. (2022)).

### 3.3. Step 2: Residuals

The second phase in identifying outlying and influential studies involves calculating residuals from the fitted models (refer to step 1 in Section 3.2). Raw residuals are computed by subtracting the predicted effect ($\hat{\mu}_i$) from the observed effect size ($y_i$). $\hat{\mu}_i$ is given by:

$$\hat{\mu}_i = \boldsymbol{x}_i\widehat{\boldsymbol{\beta}} = \hat{\beta}_0 + \hat{\beta}_1 x_{i1} + \cdots + \hat{\beta}_p x_{ip}, (13)$$

The estimated model coefficients ($\widehat{\boldsymbol{\beta}}$ or $\hat{\beta}_0, \hat{\beta}_1, \dots, \hat{\beta}_p$)) be derived using estimators such as the REML estimator outlined in Equation (9). For a common-effect (fixed-effect) model, $\hat{\theta}_i$denotes the estimated study-specific effect, whereas in an intercept-only random-effects model the fitted value reduces to the estimated overall mean effect $\hat{\mu}$. Under ideal conditions, where no studies are outliers, raw residuals should exhibit random dispersion. However, the presence of sampling variance and between-study variance may distort the residual patterns, rendering raw residuals less informative. To address this issue, Viechtbauer and Cheung (2010) proposed three standardized residual measures, drawing from the concept of standardized residuals in normal linear regression (Ling (1984), Hedges and Olkin (1985)). These measures standardize raw residuals using different denominators:

The first is Pearson (or semi-standardized) residual, standardizing raw residuals using the square root of the model-implied marginal variance (the diagonal elements of matrix $\boldsymbol{M}$ in Eqs. 6 and 7), and is expressed as:

$$r_i = \frac{(y_i - \hat{\mu}_i)}{\sqrt{v_i + \hat{\tau}^2}}, (14)$$

where $\hat{\tau}^2$ is s set to zero when computing for $r_i$ fixed-effect models, in accordance with the homogeneity assumption discussed in Section 3.1.

The second is the internally standardized residual, using the square root of the sampling variance of the raw residual as the standardizer:

$$\hat{r}_i = \frac{(y_i - \hat{\mu}_i)}{\sqrt{\mathrm{Var}[y_i - \hat{\mu}_i]}}, (15)$$

where $\mathrm{Var}[y_i - \hat{\mu}_i]$ is estimated as $(1 - h_i)(v_i + \hat{\tau}^2)$, with $h_i$ representing the leverage or hat value or "leverage" (see Section 3.4).

The third is the externally standardized residual, computed as:

$$r_{i(-1)} = \frac{(y_i - \hat{\mu}_{i(-1)})}{\sqrt{\mathrm{Var}[y_i - \hat{\mu}_{i(-1)}]}}, (16)$$

where $\hat{\mu}_{i(-1)}$ is the predicted effect size for $i$-th study after leaving out it from the model (i.e., the model is fit using the remaining $k - 1$ studies). The term $(y_i - \hat{\mu}_{i(-1)})$ is the "deleted residual", and $\mathrm{Var}[y_i - \hat{\mu}_{i(-1)}]$ is the sampling variance associated with this deleted residual "deleted residual", calculated as $\mathrm{Var}[y_i] + \mathrm{Var}[\hat{\mu}_{i(-1)}] - 2\mathrm{Cov}[y_i, \hat{\mu}_{i(-1)}]$. Given the independence between $y_i$ and $\hat{\mu}_{i(-1)}$, this simplifies to $Var[y_i] + \mathrm{Var}[\hat{\mu}_{i(-1)}] = v_i + \hat{\tau}^2_{(-1)} + \mathrm{Var}[\hat{\mu}_{i(-1)}]$, where $\hat{\tau}^2_{(-1)}$ represents the (residual) between-study variance and $\mathrm{Var}[\hat{\mu}_{i(-1)}]$ is the sampling variance of $\hat{\mu}_{i(-1)}$ after leaving out the $i$-th study from model.

Among these residual measures, the first one (Equation 14) neglects the uncertainty in the predicted effect size and does not have unit variance. While the second measure accounts for the leverage of the effect size estimate, the third measure possesses desirable statistical properties. When studies conform to the data-generation model specified in Equation 4, the third measure follows a standard normal distribution, making values exceeding 1.96 indicative of potential outliers. The terminology from normal linear regression refers to the second measure as the internally studentized residual and the third measure as the externally studentized residual. Although the externally studentized residual is widely used, two limitations should be recognized. First, its calculation relies on plug-in estimates of the within-study variance and between-study variance, and the uncertainty in these estimated quantities is not fully reflected in the resulting statistic. Second, when a conventional two-sided 5% threshold is applied to each study individually, a study has approximately a 5% probability of being flagged as unusual under the null model. Consequently, in a meta-analysis containing many studies, the probability that at least one study will exceed the threshold by chance alone can become substantial. These limitations underscore the importance of examining the characteristics of an identified outlying study rather than simply excluding it from the model based solely on the externally standardized residual.

### 3.4. Step 3: Influence measures

The third step in identifying outlying and influential studies involves calculating influence measures from the fitted models. Two key measures—weight and leverage—are particularly useful in assessing the extent to which an effect size estimate may impact the overall results.

The weight values, which are the diagonal elements of the weighting matrix $\boldsymbol{W}$, are given by the algebraic expression:

$$w_i = \frac{1}{v_i + \hat{\tau}^2}, (17)$$

The relative weight is then expressed as $w_i / \sum w_i$.

The leverage (or hat value), denoted by $h_i$, is derived from the diagonal elements of the hat matrix $\boldsymbol{H}$, which is defined as:

$$\boldsymbol{H} = \boldsymbol{X}(\boldsymbol{X}'\boldsymbol{W}\boldsymbol{X})^{-1}\boldsymbol{X}'\boldsymbol{W}, (18)$$

Unlike the weight $w_i$, the leverage value ($h_i$), incorporates both the weight and the values of the predictor variables $\boldsymbol{X}$. For random-effects models, the leverage ($h_i$) simplifies to the weight ($w_i$; Eq. 17). However, in a meta-regression model, using leverage ($h_i$) is essential since the predictor variables themselves can exert large influence on the results. For example, an extreme value in the duration of an intervention might have a disproportionate influence compared to studies with typical durations. This observation suggests that when a study is identified as influential, it should not be excluded outright; rather, the study's characteristics should be carefully examined. A leverage value greater than $3\frac{p}{k}$ indicates an influential study (Viechtbauer and Cheung (2010)).

### 3.5. Step 4: Cook's distance

The fourth step in detecting outlying and influential studies involves calculating Cook's distance. Cook's distance quantifies the impact of an individual study on the predicted effect size. Technically, it measures the Mahalanobis distance between the predicted effect size ($\hat{\mu}_i$; see Eq. 16) based on the model that includes all $k$ studies, and the predicted effect size ($\hat{\mu}_{i(-1)}$) derived from a model that excludes the $i$-th study. Drawing from the concept of Cook's distance in ordinary linear regression (Cook and Weisberg (1982)), Viechtbauer and Cheung (2010) adapted this measure for meta-analysis, defining it as:

$$D_i = \sum \frac{(\hat{\mu}_i - \hat{\mu}_{i(-1)})^2}{v_i + \hat{\tau}^2}, (19)$$

A study may be considered influential if the lower tail area of a $\chi^2$ distribution with $p$ degrees of freedom, as defined by Cook's distance, exceeds 50% (Cook and Weisberg (1982), Viechtbauer and Cheung (2010)).

### 3.6. Step 5: Changes in model coefficients and its variance

Influential studies are anticipated to impact both the model coefficients and their associated variances, thereby affecting the overall precision of the results. The fifth step in the detection process focuses on identifying influential studies by examining changes in the model coefficients and their variances. One approach involves assessing the mean difference between the predicted effect size from the full model, which includes all $k$ studies, and the predicted effect size ($\hat{\mu}_{i(-1)}$) from a model that omits the $i$-th study (Belsley et al. (2005), Viechtbauer and Cheung (2010)):

$$\text{DFFITS}_i = \frac{\hat{\mu}_i - \hat{\mu}_{i(-1)}}{\sqrt{h_i(v_i + \hat{\tau}^2_{(-1)})}}, (20)$$

$\text{DFFITS}_i$ measures how many standard deviations the predicted effect size the $i$-th study changes after excluding it during the model fitting stage. If the absolute value of $\text{DFFITS}_i$ is above $3 \times \sqrt{p(k-p)}$, it is considered as an influential study (Belsley et al. (2005), Viechtbauer and Cheung (2010)).

Additionally, the covariance ratio is used to measure the deviation in the precision of the model coefficients caused by influential studies. This metric is calculated as follows (Viechtbauer and Cheung (2010)):

$$\text{COVRATIO}_i = \frac{\text{Var}[\hat{\mu}_{i(-1)}]}{\text{Var}[\hat{\mu}_i]}, (21)$$

Since lower sampling variance corresponds to higher precision, $\text{COVRATIO}_i$ value below one suggests that excluding the $i$-th study improves the precision of the model.

### 3.7. Step 6: Model estimates of leave-one-out analysis

Remarkable alterations in any model estimates upon the exclusion of a single study can serve as indicators of outlying or influential studies (Figure 1). These estimates include the model coefficients (i.e., $\hat{\beta}_{i(-1)}$), the standard error of the model coefficients (i.e., $\text{SE}[\hat{\beta}_{i(-1)}]$), the hypothesis test related to the model coefficients (e.g., test statistic and p-value), the Cochran's Q statistic, between-study variance estimate ($\hat{\tau}^2_{i(-1)}$), and heterogeneity ($I^2_{i(-1)}$), where the subscript $i(-1)$ denotes that the $i$-th has been omitted during model fitting. Additionally, in the context of meta-regression, the pseudo-$R^2$ statistic can also be used to identify the presence of outlying and influential studies.

### 3.8. Step 7: Visual detection

Several of the measures discussed above employ "hard" cut-offs to determine whether a particular study is outlying or influential (Viechtbauer and Cheung (2010)). However, these cut-offs are somewhat arbitrary (Viechtbauer (2020)), and no single measure is entirely reliable in detecting outlying and influential studies. As the adage goes, "a picture is worth a thousand words" (Anzures-Cabrera and Higgins (2010)).

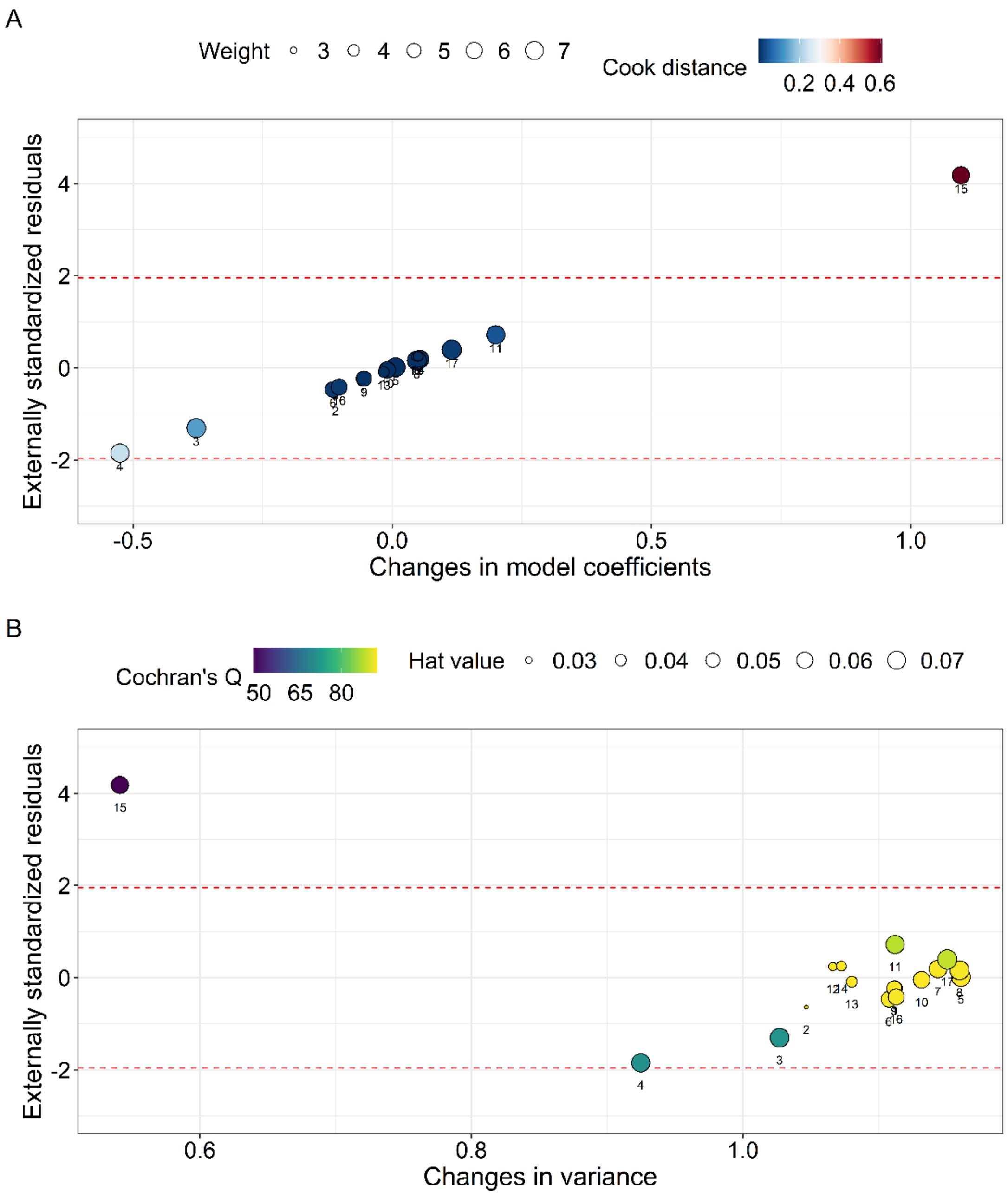

Figure 3. Visualisation solutions for detecting outlier and influential studies. (A) Bubble plots with the change in model coefficients (Eq. 20) after the exclusion of a study from the model as the *x*-axis and the externally standardized residuals (Eq. 16) as the *y*-axis. Bubble sizes indicate the weight of each study (Eq. 17), with larger weights indicating greater influence on the model coefficient estimates. Colours indicate the Cook's distance (Eq. 19). (B) An alternative bubble plot, with the variance ratio after excluding one study from the model as the *x*-axis and the external standardised residuals as the *y*-axis. Bubble size indicates the hat value (leverage; Eq. 18) assigned to each study, with larger hat values indicating a greater impact on the model coefficient estimates. Colours indicate Cochran's Q-test statistics.

Graphical methods offer a user-friendly way to visualize and identify outlying and influential studies by simultaneously displaying the values of multiple measures discussed in Sections 3.2 to 3.7. We provide two example graphical tools (Figure 3A and 3B) based on the dataset from Xiao et al. (2024)'s case study. Typically, these plots represent the externally standardized residual $r_{i(-1)}$ on the *y*-axis and the changes in model parameter estimates (e.g., model coefficients, variance) on the *x*-axis. The size of the data points is designed to be proportional to the influence measures (i.e., weight and leverage values), and a gradient colour scale is used to indicate the values of other measures, such as Cook's distance, discussed in Sections 3.2 to 3.7.

**4. Recent advances in outlying and influential studies detection techniques**

We classify recently developed techniques for detecting outlying and influential studies into three broad categories: (1) graphical methods, (2) algorithm-based methods, and (3) model-based methods. In addition, we briefly summarize approaches that account for potential outliers in meta-analysis.

### 4.1. Graphical methods

Forest and funnel plots are the most basic graphical methods for identifying outlying and influential studies. However, more sophisticated plots have been developed (Kossmeier et al. (2020), Viechtbauer (2020). For example, the Baujat plot visually detects outliers by plotting the change in model coefficients when systematically leaving out one study at a time against the contribution of this study to the Cochran's Q statistic (Baujat et al. (2002)). The GOSH (Graphical Display of Study Heterogeneity) plot, initially proposed for exploring heterogeneity, can also be used to detect outlying and influential studies (Olkin et al. (2012)). The GOSH plot visualizes the heterogeneity measure $I^2$ for all $2^k - 1$ possible combinations of $k$ studies. Another useful method is the forward plot, which monitors the effect on model parameter estimates by iteratively adding individual studies to increasingly heterogeneous sets, thereby aiding in visually detecting outliers (Mavridis et al. (2017)). In the context of meta-analysis of diagnostic test accuracy, graphical methods such as bivariate box plots (Zhou et al. (2016), Pormohammad et al. (2017)) and scatter plots of test sensitivity and specificity (Kriston et al. (2008), De Jesus et al. (2009)) have also been proposed.

### 4.2. Algorithm-based methods

The forward search (FS) algorithm is the main method developed to detect outlying and influential studies in meta-analysis. Originally designed for linear regression models (Atkinson (1994)), factor analysis (Mavridis and Moustaki (2008)), and item response theory models (Mavridis and Moustaki (2009)), the FS algorithm begins by fitting the assumed model to a small, likely outlier-free subset of studies and then adds the remaining studies based on their proximity to the postulated data-generating process. Throughout the iterations, model parameter estimates, model fit statistics (e.g., residuals, likelihood contributions), and test statistics are monitored, with sharp changes indicating potential outlying studies. This method has been extended to meta-regression (Mavridis et al. (2017)) and network meta-analysis models (Petropoulou et al. (2021)). The backward search (BS) algorithm has also been applied in meta-analysis, although it is prone to masking and swamping effects (Atkinson (1986), Hadi and Simonoff (1993)).

### 4.3. Model-based methods

Various statistical methods have been developed for identifying outlying and influential studies in meta-analysis. The benchmark "mean shift outlier model" for (univariate) random-effects meta-analysis, proposed by Viechtbauer and Cheung (2010), is a typical example (see Section 3). This model has since been extended to bivariate random-effects meta-analysis, and a likelihood ratio test (LRT) has been developed to test for the presence of outliers (Negeri and Beyene (2020)). Gumedze and Jackson (2011) proposed a "variance shift outlier model" for (univariate) random-effects meta-analysis. In network meta-analysis, identifying outlying and influential studies is crucial because such studies can reduce "consistency," a key assumption of network meta-analysis (Lu and Ades (2006)). Significant developments have been made in this area. For example, within the frequentist framework, the LRT of outlying studies has been extended to network meta-analysis (or more precisely, multivariate random-effects meta-regression (Noma et al. (2020)). In the Bayesian framework, Bayes factors based on the mean-shifted outlier model have been derived (Metelli et al. (2023)). Additionally, several analogies for outlier detection measures have been developed for generalized hierarchical models (Zhao et al. (2017)) and Bayesian hierarchical network meta-analysis models (Zhang et al. (2015)).

### 4.4. Approaches accommodating outlying studies

When outlying studies are present, we recommend more than simply excluding them from the meta-analysis. Beyond scrutinizing the characteristics of the identified outliers, a straightforward approach is to conduct the meta-analysis both with and without the outliers as part of a sensitivity analysis. More advanced approaches have also been suggested to accommodate outlying studies. One effective strategy is to adopt statistical models with flexible distributions that can tolerate outlying studies. For example, long-tailed distributions for the underlying true effects have been proposed to de-emphasize the contribution of outlying studies to the model (Baker and Jackson (2008)). Marginal distributions with additional parameters to model skewness and heavier tails have also been suggested to downweigh outlying studies (Baker and Jackson (2016)). The finite mixture method, which assumes a mixture of normal and outlying studies, is another effective approach (Beath (2014)). Another strategy is to develop robust statistical inference methods that can account for the presence of outliers. Advances have been made in such robust inference methods for meta-analysis (Basu et al. (1998)). The generalized likelihoods based on the density power divergence have been extended to meta-analysis (Noma et al. (2024)). This approach provides a statistical inference framework that is robust against outlying studies (Basu et al. (1998)).

## 5. Conclusion

In this paper, we have highlighted major issues in Xiao et al. (2024), and provided a summary of the standard practices for detecting outlying and influential studies along with three improvements. Then, we discussed the latest advancements in these techniques. It is crucial to emphasize that outlier studies and non-replicable studies represent related but conceptually different phenomena, and in some cases, a study may satisfy both definitions. We encourage the simultaneous use of techniques to detect non-replicable studies and the standard practices for identifying outlying and influential studies. When a study is identified as both non-replicable and outlying/influential, one should take the following four steps. First, verify whether the study has errors in data extraction or effect size computation (Viechtbauer (2020)). Second, assess whether including or excluding this study alters the overall conclusions of the meta-analysis. Advanced models that can account for outlying studies should also be considered to test the robustness of the meta-analysis conclusions. Next, examine whether the study has unusual characteristics, such as differences in population, intervention, or outcomes. Finally, search the existing literature or theories to locate the predictions and hypotheses relevant to the unusual study characteristics and check whether they align with the extreme effect observed in the non-replicable and outlying/influential study. By following these steps, researchers can ensure a more rigorous and reliable interpretation of meta-analytic findings.

## Author contributions

Yefeng Yang: Conceptualization; formal analysis; investigation; methodology; software; visualization; writing – original draft; writing – review and editing. SN: Conceptualization; investigation; methodology; writing – review and editing; funding acquisition; supervision. All authors approved the final manuscript.

## Competing interests

All authors declare no competing interests.

## Data and materials availability

The data and scripts needed to reproduce the analyses and figures are archived GitHub repository https://github.com/Yefeng0920/AOAS_commentary.